\documentstyle[12pt]{article}
\def\b{\begin{equation}} \def\e{\end{equation}}
\def\bd{\begin{displaystyle}} \def\ed{\end{displaystyle}}
\def\ba{\begin{array}} \def\ea{\end{array}}

\def\bee{\begin{enumerate}}
\def\eee{\end{enumerate}}

\def\bes{\begin{eqnarray*}}
\def\ees{\end{eqnarray*}}
\def\be{\begin{eqnarray}}
\def\ee{\end{eqnarray}}

\def\1{\mbox{I\hspace{-.15em}1}}

\def\b{\begin{equation}}
\def\e{\end{equation}}
\def\bee{\begin{enumerate}}
\def\eee{\end{enumerate}}

\begin{document}

\title{Krein Regularization of QED}

\author{B. Forghan\thanks{e-mail:
b.forghan@piau.ac.ir}, M.V. Takook and A. Zarei}

\maketitle  \centerline{\it Department of Physics, Science
and Research Branch, } \centerline{\it Islamic Azad University,
Tehran, Iran}

\begin{abstract}

In this paper the electron self-energy, photon self-energy and
vertex functions are explicitly calculated in Krein space
quantization including quantum metric fluctuation. The results are
automatically regularized or finite. The magnetic anomaly and Lamb
shift are also calculated in the one loop approximation in this
method. Finally, our results are compared to conventional
QED results.
\end{abstract}

\vspace{0.5cm} {\it Keywords}: Krein regularization; Krein
Quantization; Quantum metric fluctuation

\vspace{0.5cm} {\it Proposed PACS numbers}: 04.62.+v, 03.70+k,
11.10.Cd, 98.80.H \vspace{0.5cm}

\section{Introduction}

Recent observational data indicates that in the first
approximation, background space-time is very similar to de Sitter
space-time. As a result, a quantization of the linear
gravitational field without infrared divergence in such a space
time model may  be of great importance for further developments in
quantum gravity. Antoniadis, Iliopoulos and Tomaras \cite{ait}
have shown that the pathological large-distance behaviour or
infrared divergence of the graviton propagator in de Sitter
background does not manifest itself in the quadratic part of the
effective action in the one-loop approximation. This pathological behaviour of the graviton
propagator may be gauge dependent and should not appear
in an effective manner as a physical quantity. Linear gravity (the
traceless rank-2 ``massless'' tensor field) in de Sitter space is
indeed built up from copies of the minimally coupled scalar field
\cite{tt,ggt} and the above-mentioned problem also appears in the
quantization of this field. One can construct a covariant
quantization of the ``massless'' minimally coupled scalar field in
de Sitter space-time, which is causal and free of any infrared
divergence \cite{dbre, gareta1}. The essential point of this
construction is the unavoidable presence of  negative norm states,
{\it i.e.} Krein space quantization. The indefinite metric
quantization or Krein space quantization is of interest to
physicists for different reasons. The idea was first proposed by
Dirac \cite{dir} and was later exploited by Gupta and Bleuler to
remove infrared divergence in QED \cite{gup,bleu}.

The negative-norm states necessarily appear in a covariant
quantization of the free minimally coupled scalar field in de
Sitter space-time \cite{dbre,gareta1}. In this process,
ultraviolet divergence in vacuum energy and infrared divergence in
two-point function have been automatically eliminated
\cite{gareta1,ta3} and therefore  a linear graviton propagator
without infrared divergence can be obtained. In order to preserve
the covariant and eliminate the infrared divergence of linear
gravity in de Sitter space, it is necessary to use  Krein space
quantization \cite{tt,ggt,GGT}.

In the interaction QFT, the divergences appear due to the singular
behavior of the Green function at short relative distances
(ultraviolet divergence) or at large relative distances (infrared
divergence). The ultraviolet divergence appears in the following
form in the Green function, $G(x,x')$, in the limit $ x\rightarrow
x'$ or $\sigma\rightarrow 0$:
$$\frac{1}{\sigma}, \;\;\ln \sigma, \;\; \mbox{ and } \;\; \delta
(\sigma), \mbox{    where    } \sigma=\frac{1}{2}(x-x')^2. $$

It was conjectured that quantum metric fluctuations might smear
out the singularities of Green functions on the light cone {\it
i.e.} $\delta (\sigma)$, but it does not remove other ultraviolet
divergences of quantum field theory \cite{for2}. However, we have
shown that quantization in Krein space removes all ultraviolet
divergences of QFT except the light cone singularity
\cite{gareta1, ta4}. In this procedure, the auxiliary negative
norm states (negative frequency solutions which do not interact
with the physical states or real physical world) have been
utilized. It has been shown that the combination of quantum field
theory in Krein space together with the consideration of quantum
metric fluctuation results in a quantum field theory without any
divergences \cite{rota}.

A natural regularization of the one-loop interacting quantum
scalar field in Minkowski space-time $(\lambda \phi^4)$ has been
achieved through the consideration of Krein space quantization
\cite{ta4}. The Casimir effect and one-loop approximation of
Moller scattering have been determined in Krein space
\cite{khnrt,pay}. This quantization has also been developed
further to remove infrared divergence in linear gravity in de
Sitter space \cite{GGT}. Recently, the
effective action for $(\lambda \phi^4)$ theory and QED has been
calculated in Krein space quantization including quantum metric
fluctuation \cite{rerota, rerota2}. These have proven that the
physical results do not change in this method.

Although negative norm states appear in our method, by imposing
the two conditions stated below, the negative norm states
completely disappear and the theory becomes unitary:

\begin{itemize}
\item[i)] The first condition is the "reality condition" in which
the negative norm states do not appear in the external legs of the
Feynmann diagram. This condition guarantees that the negative norm
states only appear in the internal legs and in the disconnected
parts of the diagram. \item[ii)]The second condition is that the S
matrix elements must be renormalized in the following form:
$$S_{if}'\equiv\mbox{probability amplitude}=\frac{<\mbox{physical states}, in|\mbox{physical states}, out>}{<0,in|0,out>}.$$
This condition eliminates the negative norm states in the
disconnected parts.
\end{itemize}
We must emphasize the fact that this method can be used to
calculate physical observables in scenarios where the effect of
quantum gravity (in the linear approximation) can not be ignored.
In this model, the problem of non-renormalizability of linear
quantum gravity is solved.

In this paper, we explicitly calculate the photon self energy,
electron self energy and vertex function in Krein space
quantization including quantum metric fluctuation in order to
investigate the validity of our model. This method is very similar
to the Pauli-Villars regularization and then it can be considered
as a new method of regularization. Therefore we have called it
''Krein regularization''. In this method the Lagrangian does not
need any contour terms and the QFT is automatically regularized.
We also calculate the magnetic anomaly and Lamb shift in Krein
space quantization. Finally our results are compared with
conventional QED results.

\setcounter{equation}{0}
\section{Propagators in Krein space}

Let us first recall some elementary facts about the propagators in
Krein space quantization. The time-ordered product propagator for
the scalar field in Krein space quantization is \cite{gareta1}: \b
iG_T(x,x')=<0\mid T\phi(x)\phi(x') \mid 0>=\theta (t-t'){\cal
W}(x,x')+\theta (t'-t){\cal W}(x',x).\e Writing $(2.1)$ in terms
of Feynman propagator, we obtain: \b
G_T(x,x')=\frac{1}{2}[G_F(x,x')+(G_F(x,x'))^*]=\Re G_F(x,x').\e
The Feynman Green function is defined by:$$ G_F(x,x')=\int
\frac{d^4 k}{(2\pi)^4}e^{-ik.(x-x') }\tilde G_F(k)=\int \frac{d^4
k}{(2\pi)^4}\frac{e^{-ik.(x-x')}}{k^2-m^2+i\epsilon}=$$ \b
-\frac{1}{8\pi}\delta
(\sigma_0)+\frac{m^2}{8\pi}\theta(\sigma_0)\frac{J_1
(\sqrt{2m^2\sigma_0})-iN_1 (\sqrt{2m^2\sigma_0})}{\sqrt{2m^2
\sigma_0}}-\frac{im^2}{4\pi^2}\theta(-\sigma_0)\frac{K_1
(\sqrt{-2m^2\sigma_0})}{\sqrt{-2m^2 \sigma_0}},\e where
$2\sigma_0=(x-x')^2=\eta_{\mu\nu}(x^\mu-x'^\mu)(x^\nu-x'^\nu)$. So
we have: \b G_T(x,x')=\int \frac{d^4
k}{(2\pi)^4}e^{-ik.(x-x')}PP\frac{1}{k^2-m^2}=-\frac{1}{8\pi}\delta
(\sigma_0)+\frac{m^2}{8\pi}\theta(\sigma_0)\frac{J_1
(\sqrt{2m^2\sigma_0})}{\sqrt{2m^2 \sigma_0}}, \;\;x\neq x',\e
where PP stands for the principal part. This equation  exhibits
singularity on the light cone alone. The contribution of the
coincident point singularity $(x=x')$ only appears in the
imaginary part of $G_F(x,x')$  (\cite{ta3} and equation (9.52) in
\cite{bida})
$$G_F(x,x')=-\frac{2i}{(4\pi)^2}\frac{m^2}{d-4}+G^{finite}(x,x'),$$
where d is the space-time dimension and $G^{finite}(x,x')$ becomes
finite as $d\rightarrow4$.  In the momentum space for this
propagator we have \cite{itzu} \b \tilde G_T(k)=\frac{1}{2}[\tilde
G_F(k)+\tilde
G_F(k)^*]=\frac{1}{2}\left[\frac{1}{k^2-m^2+i\epsilon}+
\frac{1}{k^2-m^2-i\epsilon}\right]=PP \frac{1}{k^2-m^2}.\e

The quantum metric fluctuations
$(g_{\mu\nu}=\eta_{\mu\nu}+h_{\mu\nu})$ remove the singularities
of the Green functions on the light cone \cite{for2}. Therefore,
the quantum field theory in Krein space, including the quantum
metric fluctuation, removes all ultraviolet divergences of the
theory \cite{rota}: \b \langle G_T(x - x')\rangle = -\frac{1
}{8\pi} \sqrt{\frac{\pi}{2\langle\sigma_1^2\rangle}}
exp\left(-\frac{\sigma_0^2}{2\langle\sigma_1^2\rangle}\right)+
\frac{m^2}{8\pi}\theta(\sigma_0)\frac{J_1(\sqrt {2m^2
\sigma_0})}{\sqrt {2m^2 \sigma_0}},\e where
$2\sigma=g_{\mu\nu}(x^\mu-x'^\mu)(x^\nu-x'^\nu)$ and
$\sigma=\sigma_0+\sigma_1+...$ . In the case of $\sigma_0 =0$, due
to the metric quantum fluctuation $h_{\mu\nu}$, we have
$\langle\sigma_1^2\rangle\neq 0$, so we get \b \langle
G_T(0)\rangle = -\frac{1 }{8\pi}
\sqrt{\frac{\pi}{2\langle\sigma_1^2\rangle}} +
\frac{m^2}{8\pi}\frac{1}{2}.\e It should be noted that $
\langle\sigma_1^2\rangle $ is related to the density of gravitons
\cite{for2}.

By using the Fourier transformation of the Dirac delta function \b
-\frac{1}{8\pi}\delta(\sigma_0)=\int \frac{d^4
k}{(2\pi)^4}e^{-ik.(x-x')}PP\frac{1}{k^2}, \e for the second part
of the Green function $(2.6)$, we obtain \cite{rerota}:
\b\frac{m^2}{8\pi}\theta(\sigma_0)\frac{J_1
(\sqrt{2m^2\sigma_0})}{\sqrt{2m^2 \sigma_0}}=\int \frac{d^4
k}{(2\pi)^4}e^{-ik.(x-x')}PP\left(\frac{m^2}{k^2(k^2-m^2)}\right).\e
The first part is: \b -\frac{1 }{8\pi}
\sqrt{\frac{\pi}{2\langle\sigma_1^2\rangle}}
exp\left(-\frac{(x-x')^4}{4\langle\sigma_1^2\rangle}\right)=\int
\frac{d^4 k}{(2\pi)^4}e^{-ik.(x-x')} \tilde{G}_1(k),\e where
$\tilde{G}_1(k)$ is the Fourier transformation of the first part
of the Green function $(2.6)$. Therefore
\b <\tilde{G}_T(k)>=\tilde{G}_1(k)+PP\frac{m^2}{k^2(k^2-m^2)}.\e
In the previous paper, we showed that in the one-loop approximation,
the Green function in Krein space quantization, which appears in
the s-channel contribution of transition amplitude is \cite{ta4}:
\b  PP\frac{m^2}{k^2(k^2-m^2)}.\e It means that in this
approximation, the contribution of the first part (or quantum
metric fluctuation) is negligible. It is worth mentioning that in
order to improve the UV behavior in relativistic higher-derivative
correction theories, the propagator (2.12) has been used by some
authors \cite{ba, ho}. This propagator (2.12) also appears in the
supersymmetry theory \cite{kaku}.

The time-order product of the spinor field is constructed by its
scalar field counterpart:
\begin{equation}
\langle S_T(x-x')\rangle\equiv (i\not\partial+ m )<G_T(x,x')>,
\end{equation} where
$$iS_T(x,x')=<0\mid T \psi(x)\bar \psi(x') \mid 0>.$$ The propagator of the spinor field can be obtained as follows:
$$
\langle S_T(x-x')\rangle = \frac{1} {{8\pi }}i\mathop \gamma
\nolimits^\mu  (\mathop x\nolimits_\mu  - x'_\mu  )  \sqrt
{\frac{\pi } {{2\left\langle {\mathop \sigma \nolimits_1 ^2 }
\right\rangle }}} e^{ - \frac{{\mathop \sigma \nolimits_0 ^2 }}
{{2\left\langle {\mathop \sigma \nolimits_1 ^2 } \right\rangle }}}
\left[ { - \frac{{\mathop \sigma \nolimits_0 ^2 }} {{\left\langle
{\mathop \sigma \nolimits_1 ^2 } \right\rangle }} + \mathop
m\nolimits^2 \frac{{\mathop J\nolimits_1 (\sqrt {2\mathop
m\nolimits^2 \mathop \sigma \nolimits_0 } )}} {{\sqrt {2\mathop
m\nolimits^2 \mathop \sigma \nolimits_0 } }}} \right]+
$$
$$
\frac{1} {{8\pi }}i\mathop \gamma \nolimits^\mu  (\mathop
x\nolimits_\mu  - x'_\mu  )  \frac{m} {{2\sqrt 2 \mathop \sigma
\nolimits_0 ^{\frac{3} {2}} }}\theta (\mathop \sigma \nolimits_0 )
\left[ {\sqrt {2\mathop m\nolimits^2 \mathop \sigma \nolimits_0 }
\mathop J\nolimits_1 (\sqrt {2\mathop m\nolimits^2 \mathop \sigma
\nolimits_0 } ) - 2\mathop J\nolimits_1 (\sqrt {2\mathop
m\nolimits^2 \mathop \sigma \nolimits_0 } )} \right]+
$$
$$ \frac{m} {{8\pi }}\left[ { - \sqrt {\frac{\pi } {{2\left\langle
{\mathop \sigma \nolimits_1 ^2 } \right\rangle }}} e^{ -
\frac{{\mathop \sigma \nolimits_0 ^2 }} {{2\left\langle {\mathop
\sigma \nolimits_1 ^2 } \right\rangle }}}  + \mathop m\nolimits^2
\theta (\mathop \sigma \nolimits_0 )\frac{{\mathop J\nolimits_1
(\sqrt {2\mathop m\nolimits^2 \mathop \sigma \nolimits_0 } )}}
{{\sqrt {2\mathop m\nolimits^2 \mathop \sigma \nolimits_0 } }}}
\right] .$$ The Fourier transforation of spinor propagator is: $$
(\not
k+m)\left[\tilde{G}_1(k)+PP\frac{m^2}{k^2(k^2-m^2)}\right].$$ If
we ignore the first term the propagator becomes: \b (\not
k+m)\left[PP\frac{m^2}{k^2(k^2-m^2)}\right]. \e

The time-ordered product propagator in the Feynman gauge for the vector field in Krein space is
given by:
\begin{equation}
D_{\mu\nu}^T(x,x^\prime)=-\eta_{\mu\nu} G_T(x,x^\prime), \;\;\;
\tilde{D}_{\mu\nu}(k)=PP\frac{-\eta_{\mu\nu}}{k^2 }.
\end{equation}

It is trivial to show that in Krein space quantization, aside from
the change of propagators, the Feynman rules for QED are similar
to those of conventional QFT . In Hilbert space the propagator is
$\frac{1}{{k^2  - m^2 + i\varepsilon }}$, whereas the propagator
used in Krein space is $PP\frac{1}{{k^2  - m^2 }}$. It is
noteworthy that in tree diagrams or lines that do not appear in
the loops, these two propagators are equivalent since there are no
integrals on $k^2$ and  $k^2\neq m^2$.

For diagrams with loops there are two possibilities: i) the
diagram is convergent or ii) the diagram is divergent due to the
singularity of the delta function in the propagator. In the first
case we select the propagator to be $PP\frac{1}{{k^2  - m^2 }}$.
In the second case, in order to eliminate the delta function
singularity, the quantum metric fluctuation is included and the
propagator $(2.11)$ or alternatively an approximation of it in the
form of $(2.12)$  is used.

\setcounter{equation}{0}
\section{Electron Self Energy in Krein Space}

In this part, the electron self energy in the Krein space
quantization is calculated. It is: \b i\Sigma_{kr}(p)= ( - ie )^2
\int {\frac{{d^4 k}} {{(2\pi )^4 }}i\widetilde{D}^T_{\mu\nu}
(p-k)}{\mathop \gamma \nolimits^\mu i\widetilde{S}_T (k)\mathop
\gamma \nolimits_\mu  },\e which is divergent at the ultraviolet
limit. This divergence is due to the delta function singularity in
the propagators. By including the quantum metric fluctuation we
have \b i\Sigma_{kr}(p)= ( - ie )^2 \int {\frac{{d^4 k}} {{(2\pi
)^4 }}i<\widetilde{D}^T_{\mu\nu} (p-k)>}{\mathop \gamma
\nolimits^\mu i<\widetilde{S}_T (k)>\mathop \gamma \nolimits_\mu
},\e  which is convergent. Since we do not have the explicit form
of the Fourier transformation of $\tilde G_1$, the integral
$(3.2)$ is calculated by the following approximation: for the
electron propagator the Green function $(2.14)$ and for the photon
propagator Green function  $(2.15)$ are used. Then we obtain: \b
i\Sigma _{kr} (p)= \frac{e^2 } {4}\int \frac{d^4 k}{(2\pi )^4 }
\gamma ^\mu  (\not k + m)\gamma _\mu PP\left(\frac{1} {{k^2 - m^2
}} - \frac{1} {{k^2 }}\right)PP\left(\frac{1} {{(p - k)^2
}}\right). \e  In this approximation, the integral $(3.3)$ can be
calculated. In order to solve this integral, Feynman parameters
are used: \b i\Sigma _{kr} (p) = e^2 \int\limits_0^1 {dx} \int
{\frac{{d^4 l}} {{(2\pi )^4 }}} ( - 2x\not p + 4m)\left[\frac{1}
{{(l^2 - \Delta )^2 }} - \frac{1} {{(l^2 - \Delta ')^2 }}\right],
\e where $ k=l + xp$ , $\Delta = x^2p^2-xp^2+(1-x)m^2$ and $\Delta
' = x(x - 1)p^2 $ [Appendix A] \cite{pesc}.

The integral over $l$ is no longer divergent. By calculating the
integral over $l$, we obtain [Appendix A]:
 \b\Sigma_{kr} (p) = \frac{{e^2 }} {{8\pi ^2}}\int\limits_0^1 {dx} (x\not p - 2m)\ln \frac{{x(x - 1)p^2
 +
(1 - x)m^2 }} {{x(x - 1)p^2 }}.\e Solving the integral over $x$,
we have the following result:
$$
\Sigma _{kr} (p) = \frac{e^2 }{8\pi ^2 }\left\{ \ln \left( -
\frac{p^2 }{m^2} \right)\left(2m - \frac{\not p}{2} \right)
 \right.$$\b
\left. - \frac{\not p}{2}\left[\left(\frac{m^2 }{p^2 }\right)+
\frac{m^4  - (p^2 )^2}{(p^2 )^2 }\ln \left( 1 - \frac{p^2 }{m^2 }
\right) \right]+ 2m\left[ {\frac{m^2  - p^2 }{p^2 }\ln \left( 1 -
\frac{p^2 }{m^2} \right)} \right] \right\}.\e The result in the
Hilbert space is \cite{itzu}:
$$
   \Sigma _{Hi} (p,\Lambda ,M = 0) = \frac{{e^2 }}
{{8\pi ^2 }}\left\{ {\ln \left({\frac{{\Lambda ^2 }} {{m^2
}}}\right)\left( {2m - \frac{{\not p}} {2}} \right) + \left( {2m -
\frac{{3\not p}} {4}} \right)}  \right.$$
 \b
\left. - \frac{\not p}{2}\left[\left(\frac{m^2 }{p^2 }\right)+
\frac{m^4  - (p^2 )^2}{(p^2 )^2 }\ln \left( 1 - \frac{p^2 }{m^2 }
\right) \right]+ 2m\left[ {\frac{m^2  - p^2 }{p^2 }\ln \left( 1 -
\frac{p^2 }{m^2} \right)} \right] \right\}.\e It is clear that the
result is finite and also the last two terms  of $(3.6)$ and
$(3.7)$ are equal.

\setcounter{equation}{0}
\section{Photon Self Energy in Krein space}

The photon self energy in the Krein
space quantization is: \b i\Pi^{kr}_{\mu\nu} (k)=(-1)( -
ie)^2 \int {\frac{{d^4 p}} {{(2\pi )^4 }}Tr\left[ {\mathop \gamma
\nolimits_\mu \widetilde{S}_T (p)\mathop \gamma \nolimits_\nu
\widetilde{S}_T (p-k)} \right]}, \e which is
divergent. By including the quantum
metric fluctuation, we have \b i\Pi^{kr}_{\mu\nu} (k)=( - ie)^2
\int {\frac{{d^4 p}} {{(2\pi )^4 }}( - 1)Tr\left[ {\mathop \gamma
\nolimits_\mu <\widetilde{S}_T (p)>\mathop \gamma \nolimits_\nu
<\widetilde{S}_T (p-k)>} \right]}, \e  which is
convergent. By using the Green function $(2.14)$, we obtain:
$$
  i\Pi _{\mu \nu }^{kr} (k^2 ) = -\frac{ e^2 }
{4}\int \frac{d^4 p}{(2\pi )^4 }Tr\left[\gamma ^\mu (\not p +
m)PP\left(\frac{1} {p^2  - m^2 } - \frac{1}{p^2 }\right) \hfill
\right. \\ $$\b \left.
  \gamma ^\nu  (\not k +\not p + m)PP\left(\frac{1}
{(p + k)^2  - m^2 } - \frac{1}{(p + k)^2 }\right)\right]. \e After
using the Feynman parameter, we change the variable $p$ to $l -
xk$, perform the Wick rotation in a manner that if $i\epsilon$
exists in the denominator, the substitution $l_E^0 = - il^0$ is
used, and if $-i\epsilon$ is present in the denominator, the
change of variable $l_E^0 = il^0$ is applied. We have [Appendix B]
\cite{pesc} :
$$
  i\Pi _{\mu \nu }^{kr}  =-4ie^2 \int\limits_0^1 {dx} \int \frac{d^4 l_{E}}{(2\pi )^4 }\left[\frac{1}{(l_{E}^2+\Delta _1 )^2 } - \frac{1}
{(l_{E}^2+\Delta _2 )^2 } - \frac{1}
{(l_{E}^2+\Delta _3 )^2} \hfill \\
   + \frac{1}
{(l_{E}^2+\Delta _4 )^2 }\right]
$$\b
\left(\frac{1}{2}g^{\mu \nu } l_{E}^2 - 2x(1 - x)k^\mu  k^\nu   + g^{\mu \nu } (m^2  + x(1 - x)k^2 )\right) \hfill \\
.\e The photon self energy in Krein space is gauge invariant: \b
k^\mu  \Pi _{\mu \nu } ^{kr}(k^2 ) = 0,\e and  integral (4.4) can
be written in the following form [Appendix B]: \b\Pi _{\mu \nu
}^{kr} (k^2 ) = (k^2 g^{\mu \nu }  - k^\mu  k^\nu  )\Pi_{kr}(k^2
 ),\e
where
$$
  \Pi_{kr} (k^2 ) =  \frac{{e^2 }}
{{2\pi^2}}\int\limits_0^1 {dx} (1 - x)x \ln \left( - \frac{{k^2 }}
{{m^2 }}\right) + \frac{{e^2 }} {{2\pi^2}}\int\limits_0^1 {dx} (1
- x)x \ln \left(1 - x(1 - x)\frac{{k^2 }}
{{m^2 }}\right) \hfill \\
$$\b
   - \frac{{e^2 }}
{{2\pi^2}}\int\limits_0^1 {dx} (1 - x)x \ln \left(1 - x\frac{{k^2
}} {{m^2 }}\right) - \frac{{e^2 }} {{2\pi^2}}\int\limits_0^1 {dx}
(1 - x)x \ln \left(1 - (1 - x)\frac{{k^2 }}
{{m^2 }}\right). \hfill \\
   \hfill \\
\e The first integral yields $
  \frac{{e^2 }}
{{12\pi ^2 }}\ln\left( { -\frac{{k^2 }} {{m^2 }}} \right) $, and
the third and fourth integrals are equal to $$\frac{{4e^2 }} {{\pi
^2 }}\left\{ { - \frac{5} {{36}}} \right. - \frac{{m^2 }} {{3k^2
}} + \frac{{m^4 }} {{3k^4 }} + \left( {\frac{1} {6}} \right. -
\frac{{m^4 }} {{2k^4 }} + \left. {\frac{{m^6 }} {{3k^6 }}}
\right)\left. {\ln \left( {1 - \frac{{k^2 }} {{m^2 }}} \right)}
\right\}.$$ Therefore the photon self energy in Krein space can be
written as follows: $$
  \Pi_{kr} (k^2 ) =  \frac{{e^2 }}
{{12\pi ^2 }} \ln\left( {- \frac{{k^2 }} {{m^2 }}} \right) +
  \frac{{e^2 }} {{2\pi ^2 }}\int\limits_0^1 dx (1 - x)x\ln
\left( {1 - x(1 - x)\frac{{k^2 }} {{m^2 }}} \right)- $$\b
\frac{{4e^2 }} {{\pi ^2 }}\left\{ { - \frac{5} {{36}}} \right. -
\frac{{m^2 }} {{3k^2 }} + \frac{{m^4 }} {{3k^4 }} + \left(
{\frac{1} {6}} \right. - \frac{{m^4 }} {{2k^4 }} + \left.
{\frac{{m^6 }} {{3k^6 }}} \right)\left. {\ln \left( {1 -
\frac{{k^2 }} {{m^2 }}} \right)} \right\} . \e

In Hilbert space, the photon self energy is \cite{pesc}:
 \b
  \Pi_{Hi} (k^2 ) =  -\frac{{e^2 }}
{{12\pi ^2 }}\ln\left( { \frac{{\Lambda^2 }} {{m^2 }}} \right) +
\frac{{e^2 }} {{2\pi^2}}\int\limits_0^1 {dx} (1 - x)x\ln \left(1 -
x(1 - x)\frac{{k^2 }} {{m^2 }}\right).\e
The small part of the Lamb shift is calculated by using the
following limit:
 \b\mathop {\lim }\limits_{k^2 \to 0} k^2\Pi _{kr} (k^2
)=- \frac{{e^2}} {{60\pi^2 }}\frac{{k^4 }} {{m^2 }},\e which is
equal to the result in Hilbert space \cite{pesc}.

\setcounter{equation}{0}
\section{Vertex Function in Krein space}

The vertex function in the Krein space quantization is:
\b \Lambda _{kr}^\mu  (p',p)= (-ie)^2\int \frac{d^4k}{(2\pi)^4}
\widetilde{D}^T_{\nu\rho} (p-k)\gamma ^\nu (-i) \widetilde{S}_T
(k') \gamma ^\mu  i \widetilde{S}_T (k) \gamma^\rho ,\e and by
including the quantum metric fluctuation we have \b \Lambda
_{kr}^\mu  (p',p)= (-ie)^2\int \frac{d^4k}{(2\pi)^4}
<\widetilde{D}^T_{\nu\rho} (p-k)>\gamma ^\nu (-i) <\widetilde{S}_T
(k')> \gamma ^\mu  i <\widetilde{S}_T (k)> \gamma^\rho ,\e where
$k'-k=q$ and $p'-p=q$. After entering the results of the Fourier
transformation of propagators into Krein space and using similar
techniques, we have [Appendix C] \cite{pesc}: \b \Lambda_{kr} ^\mu
(p',p) =F_1^{kr} (q^2 )\gamma ^\mu   + \frac{{i\sigma ^{\mu \nu }
q_\nu }} {{2m}}F_2^{kr} (q^2 ),\e where ($M$ is the mass of the
photon): $$
  F_1^{kr} (q^2 ) = \frac{{e^2 }}
{8}\int\limits_0^1 {dz_3 } dz_2 dz_1 \delta (1 - z_1  - z_2  - z_3
)\int {} \frac{{d^4 k}} {{(2\pi )^4 }} PP\frac{1} {{(k - p)^2
-M^2}}PP\frac{1} {{k^2  - m^2 }} PP\frac{1} {{k'^2  - m^2 }} $$
\b\left[ { - \frac{1} {2}(k + z_2q - z_1p)^2  + q^2 (1 - z_3)(1 -
z_2) + m^2 (1 - 4z_1 +
z_1^2 )} \right] \hfill \\ , \hfill \\
\e
$$ F_2^{kr} (q^2 ) = \frac{{ie^2 }}
{8}\int\limits_0^1 {dz_3 } dz_2 dz_1 \delta (1 - z_1  - z_2  - z_3
)\int {} \frac{{d^4 k}}
{{(2\pi )^4 }}\left( {2m^2 z_1(1 - z_1)} \right) \hfill \\
  PP\frac{1}
{{(k - p)^2-M^2 }}  $$\b PP\frac{1} {{k^2  - m^2 }} PP\frac{1} {{k'^2  - m^2 }} \hfill \\
.\e It can be seen that the numerators in $(5.4)$ and $(5.5)$ are
the same as they are in Hilbert space. $F_2^{kr} (q^2 )$ is
convergent while $F_1^{kr} (q^2 )$ is divergent. This divergence
is due to the Dirac delta function; it is therefore possible to
remove the divergence via quantum metric fluctuation. In order to achieve
this, $PP\frac{{m^2}}{{k^2(k^2-m^2)}}$ is used as a propagator for
$F_1^{kr} (q^2 )$.

\subsection{$F_1^{kr}$ Term}

Due to the existence of divergence terms in this part, the Krein
regularization is used, which means the propagator $(2.14)$ is
used for the spinor field and $(2.15)$ is used for the photon
field:
$$
  F_1^{kr} (q^2 ) = \frac{{e^2 }}
{8}\int\limits_0^1 {dz_3 } dz_2 dz_1 \delta (1 - z_1  - z_2  - z_3
)\int {} \frac{{d^4 k}} {{(2\pi )^4 }} PP\frac{1} {{(k - p)^2
-M^2}}PP\left( {\frac{1} {{k^2  - m^2 }} }-{\frac{1} {{k^2
 }} } \right)$$
\b\left[ { - \frac{1} {2}(k + z_2q - z_1p)^2  + q^2 (1 - z_3)(1 -
z_2) + m^2 (1 - 4z_1 + z_1^2 )} \right] \hfill \\ PP\left(
{\frac{1} {{k'^2  - m^2 }} }-{\frac{1} {{k'^2}} } \right) .\e In
order to solve these integrals, we use Feynman parameters:
$$ \Delta _2  = m^2 (1 - z_1 )^2  + M^2 z_1  - q^2 z_2 z_3 , $$
$$ \Delta _3  = m^2 (1 - z_1 )^2  + M^2 z_1  - q^2 z_2 z_3  - z_3 m^2,$$
$$ \Delta _4  = m^2 (1 - z_1 )^2  + M^2 z_1  - q^2 z_2 z_3  - (z_2  + z_3 )m^2 , $$
\b   \Delta _5  = m^2 (1 - z_1 )^2  + M^2 z_1  - q^2 z_2 z_3  -
z_2 m^2.\e
By using the newly defined variables in equation
$(5.6)$, we obtain [Appendix D] \cite{kaku}: $$ F_1^{kr} (q^2 ) =
\frac{{e^2 }} {8\pi^2}\int {dz_1 dz_2 dz_3 \delta (1 - z_1  - z_2
- z_3 )} \sum\limits_{j = 2}^5 \left[\left( - ( -
1)^j \ln \Delta _j \right. \right.$$\b \left.\left. - \frac{{m^2
(1 - 4z_1 - z_1 ^2 ) + q^2 (1 - z_2 )(1 - z_3 )}} {{( - 1)^j
\Delta _j }}\right) \right] .\e In Hilbert space, $F_1(q^2 )$ is
as below \cite{kaku}:
 $$ F_1^{Hi} (q^2 ) =  \frac{{e^2 }}
{8\pi^2}\int {dz_1 dz_2 dz_3 \delta (1 - z_1 - z_2  - z_3 )}\left[
\left(- \ln\Delta _0 \right. \right.$$\b \left.\left.+
\frac{{m^2 (1 - 4z_1  - z_1 ^2 ) + q^2 (1 - z_2 )(1 - z_3 )}} {{
\Delta _0 }}\right) \right],\e where $\Delta _0  = m^2 (1 - z_1
)^2  + M ^2 z_1 - q^2 z_2 z_3
 - i\varepsilon  - (p'^2  - m^2 )^2 z_2 (1 - z_2 ) - (p^2  - m^2 )^2 z_3 (1 - z_3 )
$. Note that infrared divergence in $(5.8)$ is removed due to the
existence of the negative norm state of the propagators in the
loop and the answer is finite. After integrating, we have
[Appendix D]: \b F_{1}^{kr} (q^2 )_{q^2 \to 0} = \frac{{\alpha q^2
}} {{3\pi m^2 }}\left(\ln\frac{m} {M} - \frac{3}
{8}-\frac{{1}}{{4}}\right).\e
 In Hilbert space, we have \cite{itzu,kaku}:
 \b F_{1}^{Hi} (q^2 )_{q^2  \to 0} =-\frac{\alpha}{4\pi}\ln \frac{\Lambda^2}{m^2}
+\frac{{\alpha q^2 }} {{3\pi m^2 }}\left(\ln\frac{m} {M} -
\frac{3} {8}\right) .\e By using the value of $F_1(q^2)$ and the
photon self energy in Krein space, the value of the Lamb shift is
calculated to be $1018.19$ MHz, whereas in Hilbert space it was
$1052.1$ MHz. The experimental value of the Lamb shift has been
given as $1057.8$ MHz \cite{itzu}. This small difference between
the experimental results and those presented here may be because
of disregarding the linear quantum gravitational effect and also
the fact that we have worked in the one-loop approximation.

\subsection{$F_2^{kr}$ Term}

In the one-loop calculation, the magnetic anomaly does not have
any divergence terms, so the propagator $(2.5)$, which is gained
from Krein space quantization, is used.

For calculating the magnetic anomaly, at first we assume that
$q=0$ and $M=0$ in the equation $(5.5)$ then we change $k$ to $l +
z_1 p$. The variables $z_1$,$z_2$ and $z_3$ are Feynman parameters. Since the
imaginary parts cancel  each other out, only the real part
 remains \b F_2^{kr} (0)= i\frac{{8e^2 }}
{8}\int {\frac{{d^4 l}} {{(2\pi )^4 }}} \int\limits_0^1 {dz_3 }
dz_2 dz_1 \delta (1 - z_1  - z_2  - z_3 )m^2 z_1 (1-z_1)
\frac{8}{(l^2  - \Delta )^3 } , \e where $\Delta = (z_1 -1 )^2m^2
$. Consequently $F_2^{kr} (0)=\frac{{e^2}} {{8\pi^2}}$, which is
the same as the results in Hilbert space quantization.

In calculating $F_2^{kr}(q^2)$ with propagator $(2.5)$, the same
result as the conventional method is achieved. In the one-loop
approximation, $F_2^{kr}(q^2)$ does not have any divergence terms. 
The propagator $(2.12)$ is used when the singularity appears.
This propagator is very similar to the propagator which is used in
Pauli-Villars regularization, and it is because of this similarity
that we have called our method "Krein Regularization". In Hilbert
space, $F_2^{kr}(q^2)$ does not require regularization because it
does not have any divergence terms. In calculating $F_2^{kr}(q^2)$
by using $(2.12)$, the result obtained is similar to the Hilbert
space result when  Pauli-Villars regularization is used for
calculating $F_2(q^2)$.

Up to now, in addition to $F_1^{kr}(q^2)$, the electron and photon
self energy the following quantities have been calculated using
propagator (2.12) in Krein space quantization: the effective
action of $\lambda \phi^4$ \cite{ rerota}, the effective action of
QED \cite{ rerota2}, the transition amplitude of $\lambda \phi^4$
\cite{ sahraee}, the four-point function of $\lambda
\phi^4$\cite{f}. All of these items have divergence terms and
therefore using propagator (2.5) for computing them does not yield
finite results. The results obtained using (2.12) are in agreement
with the conventional method.

\section{Conclusion}

In this paper we explicitly calculated  the photon self energy,
electron self energy and vertex functions in the one-loop
approximation in Krein space quantization. We observed that the
results are finite and the Lagrangian does not need any contour
terms. The magnetic anomaly and Lamb shift are calculated in this
approximation. Our results are comparable to the results in
conventional QED. The magnetic anomaly is exactly the same as the
previous results and for the Lamb shift we have a very small
difference, which may be due to the omission of $\tilde{G}^1(k)$.

Consequently for QED, we see that this quantization eliminates the
singularity in the theory without changing the physical content of
the theory in the one-loop approximation, similar to our previous
work on the effective action for QED \cite{rerota2}. This method
can be easily used for linear quantum gravity in the background field
method, where the theory is automatically renormalized. This
method of quantization may be used as an alternative way for
solving the non-renormalizability of linear quantum gravity in the
background field method and is instrumental in finding a new
method of quantization, which would be compatible with  general
relativity.

It is important to note that  negative norm states can be
propagated in the theory, but by imposing the additional
conditions on the quantum state of the theory and the probability
amplitude, one can circumvent this problem and obtain physical
results for measurable quantities. The physical states or the
external legs of the Feynman diagram are all positive while the
negative states only appear in the internal line in the Feynman
diagram, which appear due to the effect of the S-matrix.
Therefore in calculating the S-matrix elements or probability
amplitudes for the physical states, negative norm states only
appear in the internal line and in the disconnected part of the
Feynman diagram. The negative norm states, which appear in the
disconnected part of the S-matrix elements, can be eliminated by
renormalizing the probability amplitudes
$$S_{fi}'=\frac{<\mbox{physical states}, in|\mbox{physical states}, out>}{<0,in|0,out>}.$$

We would like to submit that through this method, one can approximately calculate  physical observables when the effect
of quantum gravity can not be neglected. In this case, Krein space
quantization of linear quantum gravity is similar to the relativistic
spectrum of the hydrogen atom in the background field method
before the construction of QFT. This model can not provide a full
answer to  quantum gravity, since negative norms states appear in
our model and are eliminated by imposing the additional
conditions but this model can be applied to linear quantum gravity
easily \cite{tt,ggt,GGT,ta09,ta99}. By using the usual quantum
principle, it is impossible to quantize general relativity with
the two essential principles of general covariance and
 causality, since these two principles are closely related to locality but the quantum states in conventional QFT are defined
globally. For quantization of general relativity, the quantum
states or probability amplitudes must be defined such as
they are compatible with the general coordinate transformation,
{\it i.e.} the quantum principles for defining the
probability amplitude may be changed.

\vspace{0.5cm} \noindent {\bf{Acknowlegements}}: We are grateful
to J. Iliopoulos for his helpful discussions.

\begin{appendix}

\setcounter{equation}{0}
\section{More details about equations (3.5) and (3.7)}

In this appendix, we present the details of the calculations of
electron self energy in Krein space quantization. Using propagator
$(2.12)$ in equation $(3.3)$, we obtain $$
  i\Sigma ^{kr}(p) =  - \frac{{( - ie)^2 }}
{4}\int {\frac{{d^4 k}} {{(2\pi )^4 }}} \gamma ^\mu  (\not k +
m)\gamma _\mu  \left(\frac{1} {{k^2  - m^2  + i\varepsilon }} -
\frac{1} {{k^2 + i\varepsilon }} + \frac{1} {{k^2  - m^2  -
i\varepsilon }} - \frac{1} {{k^2  - i\varepsilon }}\right) \hfill \\
$$\b
 \left (\frac{1}
{{(p - k)^2  + i\varepsilon }} + \frac{1}
{{(p - k)^2  - i\varepsilon }}\right) \hfill \\
.\e The denominators in the above equation can be redefined as
below:

  \b yk^2  - ym^2  + (x + y)i\varepsilon  + xk^2  - 2xkp + xp^2  = l^2  - \Delta  + i\varepsilon ,
  \e
  \b
  yk^2  - ym^2  + ( - x + y)i\varepsilon  + xk^2  - 2xkp + xp^2  = l^2  -\Delta  + (1 - 2x)i\varepsilon
  ,\e
  \b
  yk^2  - ym^2  + (x - y)i\varepsilon  + xk^2  - 2xkp + xp^2  = l^2  - \Delta  - (1 - 2x)i\varepsilon ,
  \e
  \b
  yk^2  - ym^2  - (x + y)i\varepsilon  + xk^2  - 2xkp + xp^2  = l^2  - \Delta  - i\varepsilon ,
  \e
\b
  yk^2  + (x + y)i\varepsilon  + xk^2  - 2xkp + xp^2  = l^2  - \Delta ' + i\varepsilon
  ,\e
  \b
  yk^2  + ( - x + y)i\varepsilon  + xk^2  - 2xkp + xp^2  = l^2  - \Delta ' + (1 - 2x)i\varepsilon
  ,\e
  \b  yk^2  + (x - y)i\varepsilon  + xk^2  - 2xkp + xp^2  = l^2 - \Delta ' - (1 - 2x)i\varepsilon ,
  \e
  \b
  yk^2  - (x + y)i\varepsilon  + xk^2  - 2xkp + xp^2  = l^2  -\Delta ' - i\varepsilon .\e
By using the change of variable $k \to l+ xp $, we obtain:
$$
  i\Sigma ^{kr} (p) = \frac{{e^2 }}
{4}\int\limits_0^1 {dx} \int {\frac{{d^4 l}} {{(2\pi )^4 }}} ( -
2x\not p + 4m)\left[ {\frac{1} {{(l^2  - \Delta  + i\varepsilon
)^2 }}} \right. + \frac{1} {{(l^2  - \Delta  + (1 -
2x)i\varepsilon )^2 }} +  $$\b
  \frac{1}
{{(l^2  - \Delta  - (1 - 2x)i\varepsilon )^2 }} + \frac{1} {{(l^2
- \Delta  - i\varepsilon )^2 }} - \frac{1} {{(l^2  - \Delta ' +
i\varepsilon )^2 }} - \frac{1} {{(l^2  - \Delta ' + (1 -
2x)i\varepsilon )^2 }}\left. {\frac{{}} {{}}} \right]. \e
In order to solve
the integral over $l$, we apply the Wick rotation because in the
numerators we do not have any products of the terms $i\epsilon$
and $-i\epsilon$. Therefore the Wick rotation is used in a manner
that if $i\epsilon$ exists in the denominator, the substitution
$l_E^0 = - il^0$ is used, and if, conversely, $-i\epsilon$ is
present in the denominator, the change of variable $l_E^0 = il^0$
is applied. The result would be equations $(3.5)$ and $(3.7)$.

\setcounter{equation}{0}
\section{More details about equations (4.5) and (4.8)}

Now we provide further details on calculating the photon self
energy in Krein space using propagator $(2.14)$ for the electron
and propagator $(2.15)$ for the photon, with the purpose of
arriving at equations $(4.3)$ and $(4.7)$. The photon self energy
in Krein space quantization is: $$
  i\Pi _{\mu \nu }^{kr} (k) = \frac{{e^2 }}
{4}\int {\frac{{d^4 p}} {{(2\pi )^4 }}} Tr\left[ {\gamma ^\mu
(\not p + m)\gamma ^\nu  (\not p + \not k + m)} \right]$$ $$\times
  \left( {\frac{1}
{{p^2  - m^2  + i\varepsilon }} - \frac{1} {{p^2  + i\varepsilon
}} + \frac{1} {{p^2  - m^2  - i\varepsilon }} - \frac{1} {{p^2  -
i\varepsilon }}} \right) \hfill \\ $$ \b \times
  \left( {\frac{1}
{{(p + k)^2  - m^2  + i\varepsilon }} - \frac{1} {{(p + k)^2  +
i\varepsilon }} + \frac{1} {{(p + k)^2  - m^2  - i\varepsilon }} -
\frac{1} {{(p + k)^2  - i\varepsilon }}} \right). \e By using the
change of variable $p \to l - xk,$ and the Feynman parameters $x$
and $y$, the denominators of the above integral can be defined as
below: \b
  yp^2  + x(p + k)^2  - (x + y)m^2  + i\varepsilon (x + y) = l^2  + \Delta _1  + i\varepsilon ,
  \e
  \b
  yp^2  + x(p + k)^2  - (x + y)m^2  + i\varepsilon ( - x + y) = l^2  + \Delta _1  + (1 - 2x)i\varepsilon ,
  \e
  \b
  yp^2  + x(p + k)^2  - (x + y)m^2  + i\varepsilon (x - y) = l^2  + \Delta _1  - (1 - 2x)i\varepsilon ,
  \e
  \b
  yp^2  + x(p + k)^2  - (x + y)m^2  - i\varepsilon (x + y) = l^2  + \Delta _1  - i\varepsilon
  ,\e
  \b
  yp^2  + x(p + k)^2  - ym^2  + i\varepsilon (x + y) = l^2  + \Delta _2  + i\varepsilon
  ,\e
  \b
  yp^2  + x(p + k)^2  - ym^2  + i\varepsilon ( - x + y) = l^2  + \Delta _2  + (1 - 2x)i\varepsilon
  ,\e
  \b
  yp^2  + x(p + k)^2  - ym^2  + i\varepsilon (x - y) = l^2  + \Delta _2  - (1 - 2x)i\varepsilon,
  \e
  \b
  yp^2  + x(p + k)^2  - ym^2  - i\varepsilon (x + y) = l^2  + \Delta _2  - i\varepsilon ,
  \e
  \b
  yp^2  + x(p + k)^2  - xm^2  + i\varepsilon (x + y) = l^2  + \Delta _3  + i\varepsilon ,
  \e
  \b
  yp^2  + x(p + k)^2  - xm^2  + i\varepsilon ( - x + y) = l^2  + \Delta _3  + (1 - 2x)i\varepsilon ,
  \e
  \b
  yp^2  + x(p + k)^2  - xm^2  + i\varepsilon (x - y) = l^2  + \Delta _3  - (1 - 2x)i\varepsilon ,
  \e
  \b
  yp^2  + x(p + k)^2  - xm^2  - i\varepsilon (x + y) = l^2  + \Delta _3  - i\varepsilon ,
  \e
  \b
  yp^2  + x(p + k)^2  + i\varepsilon (x + y) = l^2  + \Delta _4  + i\varepsilon ,
  \e
  \b
  yp^2  + x(p + k)^2  + i\varepsilon ( - x + y) = l^2  + \Delta _4  + (1 - 2x)i\varepsilon ,
  \e
  \b
  yp^2  + x(p + k)^2  + i\varepsilon (x - y) = l^2  + \Delta _4  - (1 - 2x)i\varepsilon ,
  \e
  \b
  yp^2  + x(p + k)^2  - i\varepsilon (x + y) = l^2  + \Delta _4  - i\varepsilon ,\e
  $$
  \Delta _1  = x^2 k^2  - xk^2  + m^2 ,\hfill \\
  \Delta _2  = x^2 k^2  - xk^2  - (1 - x)m^2 ,\hfill \\ $$\b
  \Delta _3  = x^2 k^2  - xk^2  + xm^2 , \hfill \\
  \Delta _4  = x^2 k^2  - xk^2 . \hfill \\
\e This technique is similar to the one used in Hilbert space
\cite{pesc}. After performing the Wick rotation, we have:

   $$
   \Pi _{\mu \nu }^{kr} (k^2 )=\frac{{4e^2 }}
{4}\int\limits_0^1 {dx} \int {\frac{{d^4 l_{E}}} {{(2\pi )^4 }}}
\left( {\frac{{g^{\mu \nu } }} {2}l_{E}^2  - 2x(1 - x)k^\mu  k^\nu
+ g^{\mu \nu } (m^2  + x(1 - x)k^2 )} \right) \hfill \\ $$ $$
  \left[ {\frac{1}
{{(l_{E}^2  + \Delta _1 )^2 }} - \frac{1} {{(l_{E}^2  + \Delta _2
)^2 }} - \frac{1} {{(l_{E}^2  + \Delta _3 )^2 }} + \frac{1}
{{(l_{E}^2  + \Delta _4 )^2 }}} \right]
 =4e^2 \int\limits_0^1 {dx} \frac{1} {{(4\pi )^{\frac{d} {2}}
}}\Gamma \left( {2 - \frac{d} {2}} \right)$$ $$(k^2 g^{\mu \nu } -
k^\mu  k^\nu  )\left( {\frac{{}} {{}}} \right.\frac{1} {{\Delta
_1^{2 - \frac{d} {2}} }} - \frac{1} {{\Delta _2^{2 - \frac{d} {2}}
}} - \frac{1} {{\Delta _3^{2 - \frac{d} {2}} }} + \frac{1}
{{\Delta _4^{2 - \frac{d} {2}} }}\left. {\frac{{}} {{}}} \right) +
4e^2 \int\limits_0^1 {dx} \frac{1} {{(4\pi )^{\frac{d} {2}}
}}\Gamma \left( {2 - \frac{d} {2}} \right) $$\b  \left( {\frac{{}}
{{}}} \right.\frac{{m^2 g^{\mu \nu } x}} {{\Delta _1^{2 - \frac{d}
{2}} }} + \frac{{m^2 g^{\mu \nu } (1 - x)}} {{\Delta _2^{2 -
\frac{d} {2}} }} - \frac{{m^2 g^{\mu \nu } }} {{\Delta _3^{2 -
\frac{d} {2}} }}\left. {\frac{{}} {{}}} \right) = \frac{{8e^2 }}
{{16\pi ^2 }} \int\limits_0^1 {dx} x(1 - x)\ln \frac{{\Delta _1
\Delta _4 }} {{\Delta _2 \Delta _3 }}(k^2 g^{\mu \nu }  - k^\mu
k^\nu  ).\e Which is equal to equation $(4.6)$.

\setcounter{equation}{0}
\section{More details about equation (5.4) and (5.5)}
In this appendix, we clarify the calculation procedure for the
vertex function in Krein space quantization: $$
  \Lambda ^\mu _{kr} (p',p) = i\frac{{e^2 }}
{8}\int {\frac{{d^4 k}} {{(2\pi )^4 }}} \left(\frac{1} {{(k -
p)^2-M^2 + i\varepsilon }} +  \frac{1} {{(k - p)^2 -M^2 -
i\varepsilon }}\right)\gamma ^\nu  (\not k + m)\gamma ^\mu  (\not
k' + m)\gamma ^\rho \hfill
\\ $$ \b
  \left(\frac{1}
{{k^2  - m^2  + i\varepsilon }} + \frac{1} {{k^2  - m^2  -
i\varepsilon }}\right )\left(\frac{1} {{k'^2  - m^2  +
i\varepsilon }} + \frac{1} {{k'^2 - m^2  - i\varepsilon }}\right )
. \e
 By using the Feynman
parameters $z_{1}, z_{2}$ and $z_{3}$ for the numerator, changing
variable $k$ to $l-z_{2}q+z_{1}p$, we obtain \cite{pesc}: $$
  \Lambda _{kr}^\mu  (p,p') = \frac{{32}}
{8}ie^2 \int {} \frac{{d^4 l}} {{(2\pi )^4 }}\int\limits_0^1 {}
dz_1 dz_2 dz_3 \delta (1 - z_1  - z_2  - z_3 )\left[ {\gamma ^\mu
\left( { - \frac{{l^2 }} {2} + } \right.} \right. \hfill \\ $$ $$
  (1 - z_3 )(1 - z_2 )q^2  + m^2 (1 - 4z_1  + z_1 ^2 )\left. {\frac{{}}
{{}}} \right) + i\frac{{\sigma ^{\mu \nu } q_\nu  }} {{2m}}2m^2
z_1 (1 - z_1 )\left. {\frac{{}} {{}}} \right]\frac{1} {{(k - p)^2
- M^2 }} \hfill \\ $$\b
  \frac{1}
{{k^2  - m^2 }}\frac{1} {{k'^2  - m^2 }}= F_1^{kr} (q^2 )\gamma
^\mu   + i\frac{{\sigma ^{\mu \nu } q_\nu  }}
{{2m}}F_2^{kr} (q^2 ). \hfill \\
\e In the following appendices, the denominators are defined by
Feynman parameters.

\setcounter{equation}{0}
\section{More details about equations (5.8) and (5.10)}

In this part, we explicate the method used to arrive at equation
$(5.8)$. We change $k$ to $l$ as  $ k \to l + z_{1}p - z_{2}q$ and
  $\omega  = 1 - z_1$ ,  $z_2  = \omega \xi$ ,  $\theta^2=-\frac{q^2}{m^2}$ and  $\xi  = \frac{1}
{2} - \frac{1} {2}\frac{{\tanh\varphi }} {{\tanh\frac{\theta }
{2}}}$. Then the integral $(5.8)$ can be written as \cite{kaku}:
\b
  F_1 ^{kr}(q^2 ) = \frac{e^2 }
{{4\pi^2}}\left[ {\left(\ln \frac{M} {m} + 1\right)(\theta \coth
\theta - 1) - 2\coth \theta \int^{} {d\varphi\varphi \tanh\varphi
- \frac{\theta } {4}\tanh\frac{\theta } {2}} } \right] +
\sum\limits_{j = 3}^5 {p_j },  \hfill \\\e where
  \b p_3  = \frac{e^2 }
{{8\pi^2}}\int {d\xi \int {\omega d\omega \left( \ln \Delta _3 +
\frac{{m^2  + 2m^2 \omega  + m^2 \omega ^2  - 3m^2  + q^2 (1 -
\omega  + \omega ^2 \xi  - \omega ^2 \xi ^2 )}} {{(m^2  - q^2 \xi
(1 - \xi ))\omega ^2  + M^2 (1 - \omega ) - \omega (1 - \xi )m^2
}}\right)} } , \e
 \b p_4  = \frac{e^2 }
{{8\pi^2}}\int {d\xi \int {\omega d\omega \left(\ln \Delta _4  -
\frac{{m^2  + 2m^2 \omega  + m^2 \omega ^2  - 3m^2  + q^2 (1 -
\omega  + \omega ^2 \xi  - \omega ^2 \xi ^2 )}} {{(m^2  - q^2 \xi
(1 - \xi ))\omega ^2  + M^2 (1 - \omega ) - \omega m^2 }}\right)}
}, \e
 \b p_5  = \frac{e^2}
{{8\pi^2}}\int {d\xi \int {\omega d\omega \left(\ln \Delta _5  +
\frac{{m^2  + 2m^2 \omega  + m^2 \omega ^2  - 3m^2  + q^2 (1 -
\omega  + \omega ^2 \xi  - \omega ^2 \xi ^2 )}} {{(m^2  - q^2 \xi
(1 - \xi ))\omega ^2  + M^2 (1 - \omega ) - \omega \xi m^2
}}\right)} }. \e  The equation $(D.2)$, $(D.3)$ and $(D.4)$ can be
written in the form of $
  p_3  = \sum\limits_{i = 1}^4 {p_{i3} }$, $
  p_4  = \sum\limits_{i = 1}^4 {p_{i4} }$, $
  p_5  = \sum\limits_{i = 1}^4 {p_{i5} }$, where
  \b p_{13}  = \frac{e^2}
{{8\pi^2}}\int {d\xi \int {\omega d\omega \ln [(m^2  - q^2 \xi (1
- \xi ))\omega ^2  + M^2 (1 - \omega ) - \omega (1 - \xi )m^2 ]} }
, \e \b p_{23}  =  \frac{e^2 } {{8\pi^2}}\int {d\xi \int {\omega
d\omega \frac{{ - 2m^2  + q^2 }} {{(m^2  - q^2 \xi (1 - \xi
))\omega ^2  + M^2 (1 - \omega ) - \omega (1 - \xi )m^2 }}} } , \e
  \b p_{33}  = \frac{e^2}
{{8\pi^2 }}\int {d\xi \int {\omega d\omega \frac{{(2m^2  - q^2
)\omega }} {{(m^2  - q^2 \xi (1 - \xi) )\omega ^2  +M^2 (1 -
\omega ) - \omega (1 - \xi )m^2 }}} } , \e
  \b p_{43}  =  \frac{e^2 }
{{8\pi^2}}\int {d\xi \int {\omega d\omega \frac{{(m^2  + q^2 (\xi
(1 - \xi ))\omega ^2 }} {{(m^2  - q^2 \xi (1 - \xi ))\omega ^2  +
M ^2 (1 - \omega ) - \omega (1 - \xi )m^2 }}} } , \e
 \b p_{14}  =  - \frac{e^2}
{{8\pi^2 }}\int {d\xi \int {\omega d\omega \ln [(m^2  - q^2 \xi (1
- \xi ))\omega ^2  + M^2 (1 - \omega ) - \omega m^2 ]} } , \e
 \b  p_{24}  = -\frac{e^2}
{{8\pi^2 }}\int {d\xi \int {\omega d\omega \frac{{ - 2m^2  + q^2
}} {{(m^2  - q^2 \xi (1 - \xi ))\omega ^2  +M^2 (1 - \omega ) -
\omega m^2}}} } , \e
 \b  p_{34}  = -\frac{e^2}
{{8\pi^2 }}\int {d\xi \int {\omega d\omega \frac{{(2m^2  - q^2
)\omega }} {{(m^2  - q^2 \xi (1 - \xi ))\omega ^2  +M^2 (1 -
\omega ) - \omega m^2 }}} } , \e
 \b p_{44}  = -\frac{e^2}
{{8\pi^2 }}\int {d\xi \int {\omega d\omega \frac{{(m^2  + q^2 (\xi
(1 - \xi ))\omega ^2 }} {{(m^2  - q^2 \xi (1 - \xi ))\omega ^2  +
M^2 (1 - \omega ) - \omega m^2 }}} } , \e
  \b p_{15}  = \frac{e^2}
{{8\pi^2 }}\int {d\xi \int {\omega d\omega \ln [(m^2  - q^2 \xi (1
- \xi ))\omega ^2  +M^2 (1 - \omega ) - \omega \xi m^2 ]} } , \e
 \b  p_{25}  =   \frac{e^2}
{{8\pi^2 }}\int {d\xi \int {\omega d\omega \frac{{ - 2m^2  + q^2
}} {{(m^2  - q^2 \xi (1 - \xi ))\omega ^2  +M^2 (1 - \omega ) -
\omega \xi m^2 }}} } , \e
 \b p_{35}  =   \frac{e^2}
{{8\pi^2 }}\int {d\xi \int {\omega d\omega \frac{{(2m^2  - q^2
)\omega }} {{(m^2  - q^2 \xi (1 - \xi ))\omega ^2  + M^2 (1 -
\omega ) - \omega \xi m^2 }}} }, \e
 \b  p_{45}  =  \frac{e^2}
{{8\pi^2 }}\int {d\xi \int {\omega d\omega \frac{{(m^2  + q^2 (\xi
(1 - \xi ))\omega ^2 }} {{(m^2  - q^2 \xi (1 - \xi ))\omega ^2  +
M^2 (1 - \omega ) - \omega \xi m^2}}} } . \e
For solving these
integrals, we define the new variable $\xi$ as $ \xi  = \frac{1}
{2} - \frac{1} {2}\frac{{\tanh\varphi }} {{\tanh\frac{\theta }
{2}}}$ and $$
  d\xi  =  - \frac{1}
{{2\tanh\frac{\theta } {2}\cosh ^2 \varphi }}d\varphi. $$
Consequently:  $$ p_{23}  + p_{25}+p_{33} + p_{35}+p_{43} +
p_{45}+p_{24}+p_{34}=$$ \b \frac{\alpha } {{4\pi }}\theta^2
  -\frac{\alpha } {{3\pi }}\theta^2-\frac{2\alpha } {{9\pi }}\theta^2+\frac{\alpha } {{2\pi }}\theta^2 +\frac{\alpha }
{\pi }\theta \coth \theta tgh^2 \frac{\theta } {2}\ln \theta ^2
 , \e
   \b p_{14}  = -\frac{\alpha } {{36\pi }}\theta^2,
  p_{15}  + p_{13}  =\frac{\alpha } {{8\pi }}\theta^2+ \mbox{constant},\e
 \b
  p_{44}  =-\frac{\alpha } {{3\pi }}\theta^2-\frac{\alpha } {{8\pi }}\theta^2-\frac{\alpha } {{4\pi }}\theta^2-\frac{\alpha } {{2\pi }}\theta^2+\frac{\alpha } {{\pi }}\theta^2 +\frac{\alpha }
{{2\pi }}\ln \theta ^{^2 } .\e When $\theta^2 \to 0$ and
$\frac{{\theta}}{{\sinh\theta}}\to 1$, all $p_{ij}$ terms are
added and equation $(5.10)$ is proved.

\end{appendix}

\end{document}